# Chapter 4

# Fog Computing: Principles, Architectures, and Applications

*Amir Vahid Dastjerdi, Harshit Gupta, Rodrigo N. Calheiros, Soumya K. Ghosh, and Rajkumar Buyya*

***Abstract-*** The Internet of Everything (IoE) solutions gradually bring every object online, and processing data in centralized cloud does not scale to requirements of such environment. This is because, there are applications such as health monitoring and emergency response that require low latency and delay caused by transferring data to the cloud and then back to the application can seriously impact the performance. To this end, Fog computing has emerged, where cloud computing is extended to the edge of the network to decrease the latency and network congestion. Fog computing is a paradigm for managing a highly distributed and possibly virtualized environment that provides compute and network services between sensors and cloud data centers. This chapter provides background and motivations on emergence of Fog computing and defines its key characteristics. In addition, a reference architecture for Fog computing is presented and recent related development and applications are discussed.

***Keywords*-** *Internet of Things; IoT; Web of Things; Cloud of Things; Fog Computing; IoT Applications; Edge Computing.*

## 4.1 Introduction

IoT environments consist of loosely connected devices that are connected through heterogeneous networks. In general, the purpose of building such environments is collecting and processing data from IoT devices to mine and detect patterns, or perform predictive analysis or optimization and finally make smarter decision in a timely manner. Data in such environment can be classified to two categories [16]:

- Little Data or Big Stream: transient data that is captured constantly from IoT smart devices.

- Big data: persistent data and knowledge stored and archived in centralized cloud storage.

IoT environments including smart cities and infrastructures need both Big Stream and Data for effective real-time analytics and decision making. This can enable real time cities [17] that are capable of real-time analysis of city infrastructure and life and provides new approaches for governance. At the moment, data is collected and aggregated from IoT networks that consist of smart devices and is sent uplink to cloud servers where it is stored and processed. Cloud computing offers a solution at the infrastructure level that supports Big Data Processing. It enables highly scalable computing platforms that can be configured on demand to meet constant changes of applications requirements in a pay-per-use mode, reducing the investment necessary to build the desired analytics application. As mentioned above, this perfectly matches requirement of Big Data processing when data is stored in centralized cloud storage. In such case, processing of large magnitude of data volume is enabled by on-demand scalability of Clouds. However, when data sources are distributed across multiple locations and low latency is indispensable, in-cloud data processing fails to meet the requirements.

## 4.2   Motivation Scenario

A recent analysis [18] of Endomondo application, a popular sport activity tracking application has revealed number of remarkable observations. The study shows that a single workout generates 170 GPS tuples, and the total number of GPS tuples can reach 6.3 million in a month time. With 30 million users (as shown in Figure 4.1), the study shows that generated data flows of Endomondo can reach up to 25,000 tuple per second. Therefore, one can expect that data flows in real-time cities with many times more data sources—GPS sensors in cars to air and noise pollution sensors—can easily reach millions of tuples per second. Centralized cloud servers cannot deal with flows with such velocity in real-time. In addition, considerable numbers of users, due to privacy concerns, are not comfortable to transfer and store activity track data into the cloud even if they require statistical report on their activities. This motivates the need of alternative paradigm that is capable of bringing the computation to more computationally capable devices that are geographically closer to the sensors than to the clouds and that have connectivity to the Internet. Such devices, which are in the *edge of the network* and therefore referred to as *edge devices,* can build local views of data flows and can aggregate data to be sent to the cloud for further off-line analysis. To this end, Fog computing has emerged.

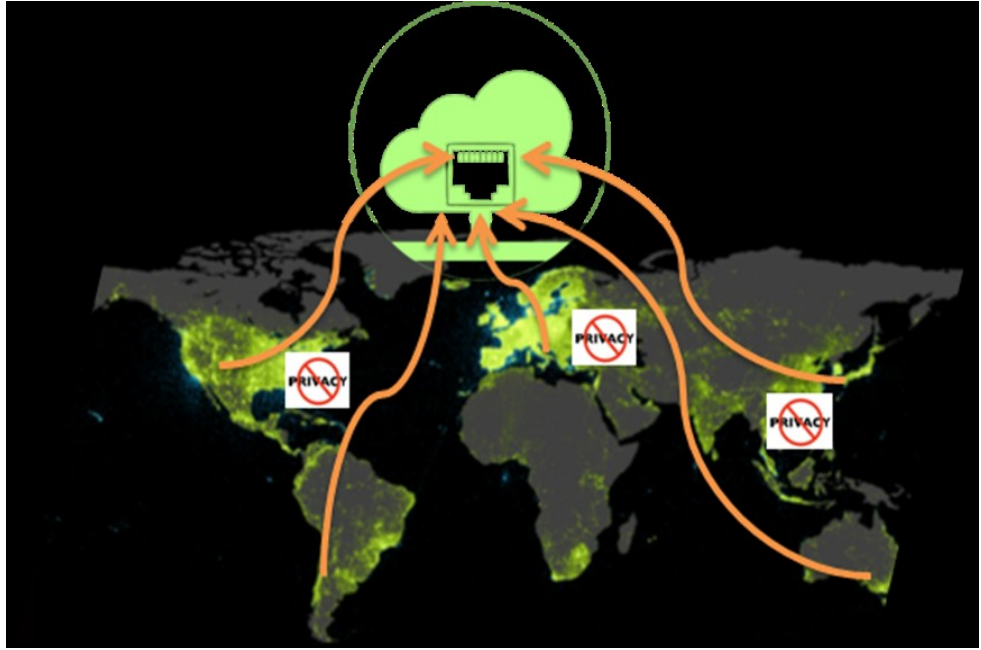

Figure 4.1. Endomondo has 30 million users around the globe generating 25000 records per second. Centralized processing of the data flow of this magnitude neither satisfies latency constraints of users nor their privacy constraints.

## 4.3   Definitions and Characteristics

We define Fog computing as a distributed computing paradigm that fundamentally extends the services provided by the cloud to the edge of the network (as shown in Figure 4.2). It facilitates management and programming of compute, networking and storage services between data centers and end devices. Fog computing essentially involves components of an application running both in the cloud as well as in edge devices between sensors and the cloud, i.e. in smart gateways, routers or dedicated fog devices. Fog computing supports mobility, computing resources, communication protocols, interface heterogeneity, cloud integration, and distributed data analytics to addresses requirements of applications that need low latency with a wide and dense geographical distribution.

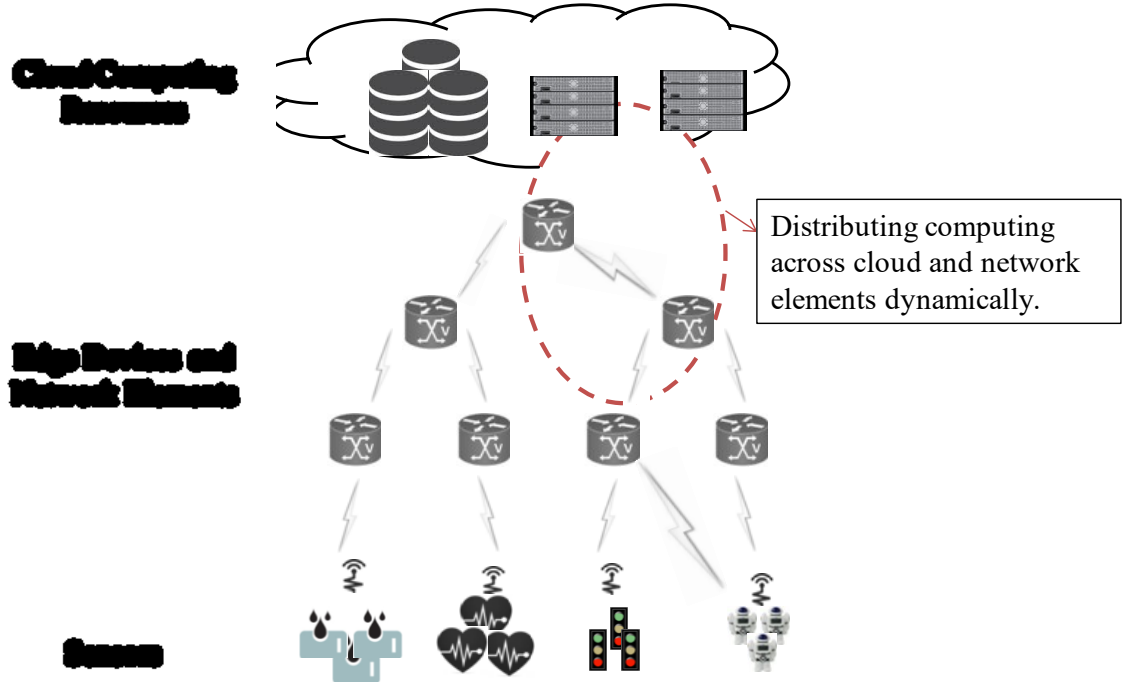

Figure 4.2 Fog computing is a distributed computing paradigm that extends the services cloud service to the edge of the network.

Advantages associated with Fog computing including the following:

- **Reduction of network traffic:** Cisco estimates that there are currently 25 billion connected devices worldwide, a number that could jump to 50 billion by 2020. The billions of mobile devices such as smart phones and tablets already being used to generate, receive and send data make a case for putting the computing capabilities closer to where devices are located, rather than having all data sent over networks to central data centers. Depending on the configured frequency, sensors may collect data every few seconds. Therefore, it is neither efficient nor sensible to send all of this raw data to the cloud. Hence, fog computing benefits here by providing a platform for filter and analysis of the data generated by these devices close to the edge, and for generation of local data views. This drastically reduces the traffic being sent to the cloud.

- **Suitable for IoT tasks and queries**: With the increasing number of smart devices, most of the requests pertain to the surroundings of the device. Hence, such requests can be served without the help of the global information present at the cloud. For example, the aforementioned sports tracker application Edomondo allows a user to locate people playing a similar sport nearby. Because of the local nature of the typical requests made by this application, it makes sense that the requests are processed in fog rather than cloud infrastructure. Another example can be a smart connected vehicle which needs to capture events only about a hundred meters from it. Fog computing makes the communication distance closer to the physical distance by bringing the processing closer to the edge of network.

- **Low latency requirement:** Mission critical applications require real-time data processing. One of the best examples of such applications is cloud robotics, control of fly-by-wire aircraft, or anti-lock brakes on a vehicle. For a robot, motion control depends on the data collected by the sensors and the feedback of the control system. Having the control system running on the cloud may make the sense-process-actuate loop slow or unavailable as result of communication failures. This is where fog computing helps by performing the processing required for control system very close to the robots - thus making real-time response possible.

- **Scalability:** Even with virtually infinite resources, the cloud may become the bottleneck if all the raw data generated by end devices is continued to be sent to it. Since fog computing aims at processing incoming data closer to the data source itself, it reduces the burden of that processing on the cloud, thus addressing the scalability issues arising out of the increasing number of endpoints.

## 4.4 Reference Architecture

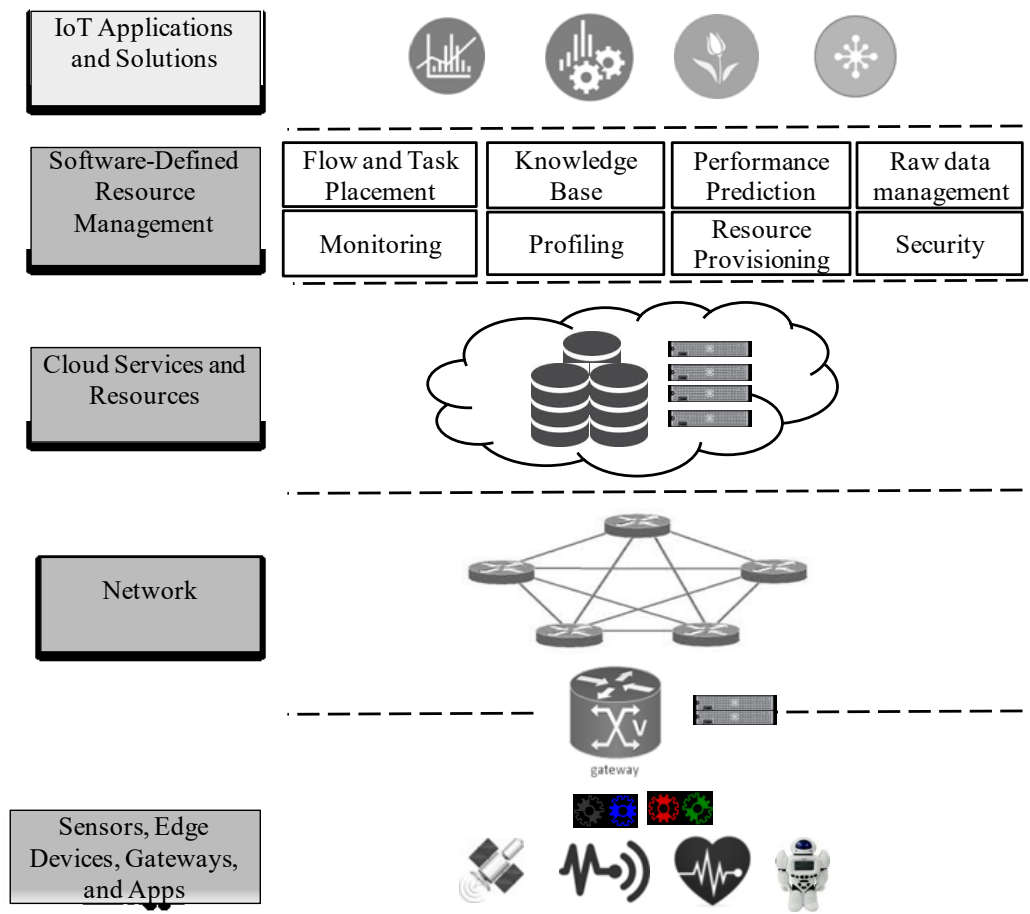

Figure 4.1. Fog computing reference architecture.

Figure 4.3 presents a reference architecture for fog computing. In the bottommost layer lie the end devices (sensors), as well as edge devices and gateways. This layer also includes apps that can be installed in the end devices to enhance their functionality. Elements from this layer use the next layer, the network, for communicating between themselves, and between them and the cloud. The next layer contains the cloud services and resources that support resource management and processing of IoT tasks that reach the cloud. On top of the cloud layer lays the resource management software that manage the whole infrastructure and enable quality of Service to Fog Computing applications. Finally, the topmost layer contains the applications that leverage fog computing to deliver innovative and intelligent applications to end users.

Looking inside the Software-Defined Resource Management layer, it implements many middleware-like services to optimize the use of the cloud and Fog resources on behalf of the applications. The goal of these services is to reduce the cost of using the cloud at the same time that performance of applications reach acceptable levels of latency by pushing task execution to Fog nodes. This achieved with a number of services working together, as follows.

- *Flow and task placement:* this component keeps track of the state of available cloud, Fog and network resources (information provided by the Monitoring service) to identify the best candidates to hold incoming tasks and flows for execution. This component communicates with the Resource Provisioning service to indicate the current number of flows and tasks, which may trigger new rounds of allocations if deemed too high.
- *Knowledge Base:* This component stores historical information about application demand and resource demands that can be leveraged by other services to support their decision-making process.
- *Performance Prediction:* This service utilizes information of the Knowledge Base service to estimate the performance of available cloud

resources. This information is used by the Resource Provisioning service to decide the amount of resources to be provisioned. In times where there is a large number of tasks and flow in use or when performance is not satisfactory.

- *Raw Data Management:* This service has direct access to the data sources and provides views from the data for other services. Sometimes, these views can be obtained by simple querying (e.g, SQL, or NOSQL REST APIs), whereas other times more complex processing may be required (e.g, MapReduce). Nevertheless, the particular method for generation of the view is abstracted away from other services.

- *Monitoring.* This service keeps track of the performance and status of applications and services and supplies this information to other services as required.

- *Profiling.* This service builds resource and applications profiles based on information obtained from the Knowledge Base and Monitoring services.

- *Resource Provisioning:* This service is responsible for acquiring cloud Fog and network resources for hosting the applications. This allocation is dynamic, as requirements of applications, as well as number of hosted applications, changes over time. Decision on the number of resources is made with use of information provided by other services (such as Profiling, Performance Prediction, and Monitoring) and user requirements on latency as well as credentials managed by the Security service. For example, the component pushes tasks with low latency requirements to edge of network as soon as free resources are available.

- *Security:* This service supplies authentication, authorization, and cryptography as required by services and applications.

Notice that all the elements and services described are reference elements only; Complete fog stacks and applications can be built without the use of all the elements, or can be built with other elements and services not present in Figure 4.3.

## 4.5 Applications

As demonstrated in Figure 4.4, there is variety of applications benefiting from Fog computing paradigm. We discuss the major applications first, and then we elaborate more on enablers and related works in the area.

**Healthcare**

Cao et al. [1] propose FAST, a fog computing assisted distributed analytics system to monitor fall for stroke patients. The authors have developed a set of fall detection algorithms, including algorithms based on acceleration measurements and time series analysis methods, as well as filtering techniques to facilitate fall detection process. They designed a real-time fall detection system based on fog computing that divides the fall detection task between edge devices and the cloud. The proposed system achieves a high sensitivity and specificity when tested against real-world data. At the same time, the response time and energy consumption are close to the most efficient existing approaches.

Another use of fog computing in healthcare has been brought out by Stantchev et al. [2]. They proposed three-tier architecture for smart-healthcare infrastructure, comprising of a role model, layered cloud architecture, and a fog-computing layer in order to provide an efficient architecture for healthcare and elderly-care applications. The fog layer improves the architecture by providing low latency, mobility support, location awareness, and security measures. The process flow of the healthcare application is modeled using Business Process Modeling Notation (BPMN) and is then mapped to devices via a service-oriented approach. The validity of architectural model has been demonstrated by a use-case as a template for a smart sensor-based healthcare infrastructure.

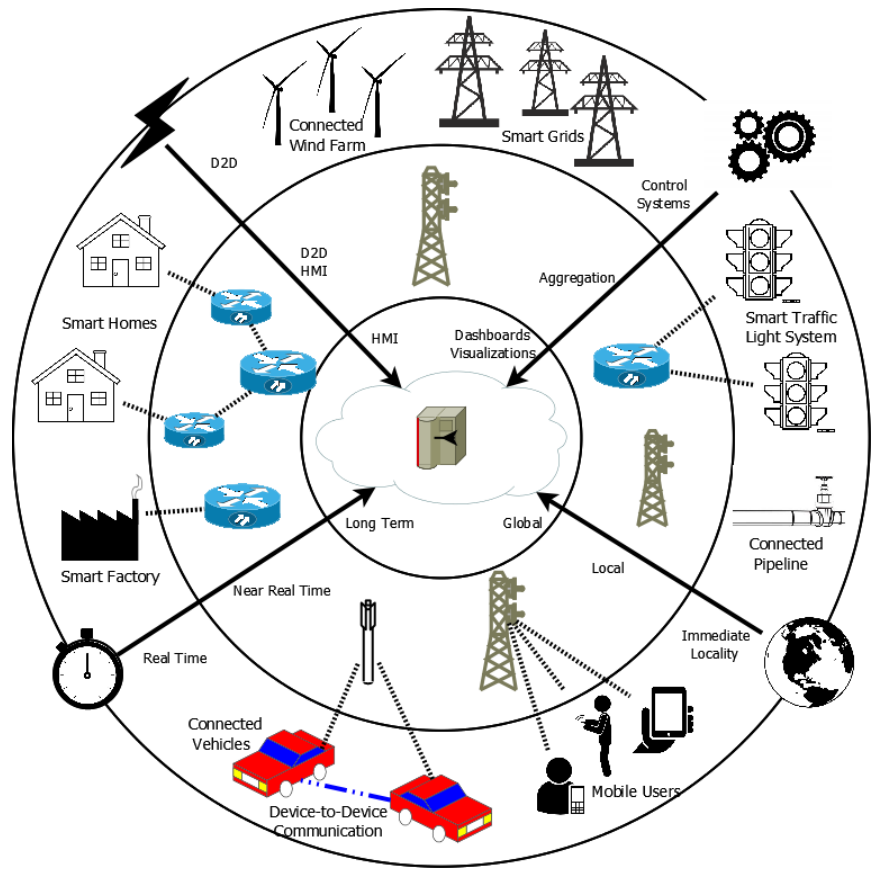

Figure 4.2 : Range of applications benefitting from Fog Computing

**Augmented Reality**

Augmented reality applications are highly latency-intolerant as even very small delays in response can damage the user experience. Hence, fog computing has the potential to become a major player in the augmented reality domain. Zao et al. [5] built an Augmented Brain Computer Interaction Game based on Fog Computing and Linked Data. When a person plays the game, raw streams of data collected by EEG sensors are generated and classified to detect the brain state of the player. Brain state classification is among the most computationally heavy signal processing tasks, but this needs to be carried out in real-time. The system employs both fog and cloud servers, a combination that enables the system to perform continuous real-time brain state classification at the fog servers while the classification models are tuned regularly in the cloud servers based on the EEG readings collected by the sensors.

Ha et al. [6] propose a Wearable Cognitive Assistance system based on Google Glass devices that assists people with reduced mental acuity. Because of the nature of cognitive devices with constrained resources, the compute-intensive workloads of this application need to be offloaded to an external server. However, this offloading must provide crisp, real-time responses, failing to do which will be detrimental to the user experience. Offloading the compute-intensive tasks to the cloud incurs a considerable latency, thus the authors make use of nearby devices. These devices may communicate with the cloud for delay-tolerant jobs like error reporting and logging. The aforementioned works are typical applications of fog computing in that they perform latency-critical analysis at the very edge and latency-tolerant computation at the cloud – thus portraying fog as an extension of cloud.

**Caching and Preprocessing**

Zhu et al. [3] discuss the use of edge servers for improving web sites performance. Users connect to the internet through fog boxes - hence each HTTP request made by a user goes through a fog device. The fog device performs a number of optimizations that reduces the amount of time the user has to wait for the requested web page to load. Apart from generic optimizations like caching HTML components, reorganizing webpage composition, and reducing size of web objects, edge devices also perform optimizations that take user behavior and network conditions into account. For example, in case of network congestion, the edge device may provide low resolution graphics to the user to reach acceptable response times. Furthermore, the edge device can also monitor the performance of the client machines and, depending on the browser rendering times, send graphics of an appropriate resolution.

One of the major uses of fog computing is how the link IoT and cloud computing. This integration is not trivial and involved several challenges. One of the most important challenges is data trimming. This trimming or pre-processing of data before sending it to the cloud will be a necessity in IoT environments because of the huge amount of data generated by these environments. Sending huge volumes of raw data to the cloud will lead to both core network and data center congestion. To meet the challenge of pre-processing, Aazam et. al [4] propose a smart gateway based communication for integrating IoT with cloud computing. Data generated by IoT devices is sent to the smart gateway, either directly (one-hop) or through sink nodes (multi-hop). The smart gateway handles the pre-processing required before sending the data to the cloud. In the architecture proposed by the authors, the smart gateway is assisted by fog computing services for operations on IoT data in a latency-sensitive and context-aware manner. Such a communication approach paves the way for the creation of richer and better user experience for IoT applications.

## 4.6 Research Directions and Enablers

To realize the full potential of Fog paradigm, researchers and practitioners need to address following major challenges.

**Programming Models**

Computation offloading has been an active area of research in the mobile computing domain, with most of the proposals offloading workloads to the cloud [6][7][8]. Since offloading to the cloud may not always be possible or reasonable, Orsini et al. [9] propose an adaptive Mobile Edge Computing (MEC) programming framework named CloudAware [9], which offloads tasks to edge devices, thus facilitating the development of elastic and scalable edge-based mobile applications. The authors present the types of components that an MEC application should be broken into, so that the offloading decision is simplified. The framework offloads tasks with the objective of one of i) speed up computation, ii) save energy, iii) save bandwidth, or iv) provide low latency.

The most fundamental development in the realm of Fog Computing has been made by Mobile Fog [10], an API for developing futuristic applications which leverage the large-scale, geo-distribution and low latency guarantee provided by the fog computing infrastructure. The proposed architecture is a hierarchical similar to the one demonstrated in Figure 4.3. An application built using the proposed API has several components, each component running on a different level in the hierarchy of devices.

**Security and Reliability**

Enforcing security protocols over a distributed system such as a fog is one of the most important challenges in its realization. Stojmenovic et al. [15] discussed the major security issues in fog computing. They pointed out that calling authentication at various levels of fog nodes is the main security challenge. Authentication solutions based on Public Key Infrastructure [16] may prove beneficial for this problem. Trusted execution environment (TEE) techniques [17,18] are potential solutions to this authentication problem in fog computing as well. Measurement-based methods may also be used to detect rogue devices and hence reduce authentication cost [18,19].

Dsouza et al. [20] describe the research challenges in policy management for fog computing and propose a policy-driven security management approach including policy analysis and its integration with fog computing paradigm. Such an approach is critical for supporting secure sharing, and data reuse in heterogeneous Fog environments. The authors also present a use-case on Smart Transportation Systems to highlight the efficiency of the proposed approach.

Since fog computing is realized by the integration of a large number of geographically distributed devices and connections, reliability is one of the prime concerns when designing such a system. Madsen et al. [11] discuss the reliability issues associated with fog computing. They pointed out that for a reliable fog paradigm it is essential to plan for failure of individual sensors, network, service platform, and the application. To this end, the current reliability protocols for WSNs can be applied. They are majorly deal with packet reliability and event reliability. The most basic facts about sensors, in general, are not expensive but their readings can be affected by noise, in this case the information accuracy problem can be solved by redundancy

**Resource Management**

Fog devices are often network devices equipped with additional storage and compute power. However, it is difficult for such devices to match the resource capacity of traditional servers, left alone the cloud. Hence a judicious management of resources is essential for an efficient operation of a fog computing environment. Aazam et al. [12] presented a service oriented resource management model for fog computing, which performs efficient and fair management of resources for IoT deployments. The proposed resource management framework predicts the resource usage of customers and pre-allocates resources based on user-behaviour and the probability of using it in the future. This prediction allows greater fairness and efficiency when the resources are actually consumed. Lewis et al. [14] present resource provisioning mechanisms for tactical cloudlets, a strategy for providing infrastructure to support computation offloading and data staging at the tactical edge. Cloudlets are discoverable, generic, stateless servers located in single-hop proximity of mobile devices, that can operate in disconnected mode and are virtual-machine (VM) based to promote flexibility, mobility, scalability, and elasticity [13]. In other words, tactical cloudlet refers to the scenario when cloudlets serve as fog devices in order to provide infrastructure to offload computation, provide forward data-staging for a mission, perform data filtering to remove unnecessary data from streams intended for dismounted users, and serve as collection points for data heading for enterprise repositories. Tasks running on cloudlets are executed on Service VMs. The authors propose various policies for provisioning VMs on cloudlets, each policy having a unique implication on payload sent to cloudlet, application-ready time and client energy spent. In addition, mechanisms for cloudlet discovery and application execution have also been laid out.

**Energy Minimization**

Since fog environments involve the deployment of a large number of fog nodes, the computation is essentially distributed and can be less energy-efficient than the centralized cloud model of computation. Hence, the reduction of energy consumption in fog computing is an important challenge. Deng et al. [15] study the trade-off between power consumption and delay in a fog computing system. They model the power consumption and delay functions for the fog system and formalize the problem of allocating workloads between the fog and cloud. Simulation results show that fog computing can significantly cut down the communication latency by incurring slightly greater energy consumption.

Do et al.[16] study a related problem, namely joint resource allocation and reduction of energy consumption for video streaming service in fog computing. Since the number of fog devices is enormous, a distributed solution for the problem has been proposed to eliminate performance and scalability issues. The algorithm is based on proximal algorithms, a powerful method for solving distributed convex optimization problems. The proposed algorithm has a fast convergence rate with a reasonable solution quality.

## 4.7 Commercial Products

**Cisco IOx**

Cisco is a pioneer in the field of fog computing, so much so, that the term fog computing was actually introduced by Cisco itself. Cisco's offering for fog computing, known as IOx, is a combination of industry-leading networking operating system, IOS and the most popular open source Operating System, Linux. Ruggedized routers running Cisco IOx make compute and storage available to applications hosted in a Guest Operating System running on a hypervisor alongside the IOS virtual machine. Cisco provides an app-store which allows users to download applications to the IOx devices and an app-management console which is meant for controlling and monitoring the performance of an application. Using device abstractions provided by Cisco IOx APIs, applications running on the fog can communicate with IoT devices that use any protocol. The "bring your own interface" philosophy of IOx allows effortless integration of novel, specialized communications technology with a common IP architecture. Fog applications can also send IoT data to the cloud by translating non-standard and proprietary protocols to IP.

Cisco IOx has been used by a number of players in the IoT industry to architect innovative solutions to problems. For example, Rockwell developed FactoryTalk AssetCentre, a centralized tool for secure tracking and management of automation related asset information across the entire plant. OSIsystem's PI system, an industry standard in enterprise infrastructure for real-time event and data management, uses Cisco IOx to deploy its data collection interfaces.

**Data In Motion**

Cisco Data In Motion (DMo) is a technology providing data management and analysis at the edge of the network. Cisco DMo is built into solutions provided by Cisco and its partners. DMo provided a simple rule-based RESTful API for building applications. Rules can be added/deleted on the run without any downtime. DMo can be use to perform analysis on incoming data such as finding specific data of interest, summarizing data, generating new result from data, etc. It is meant to be deployed on devices in a distributed fashion and control the flood of data originating from the IoT devices.

**LocalGrid**

LocalGrid's Fog Computing platform is an embedded software installed on network devices (switches, routers) and sensors. It standardizes and secures communications between all kinds of devices across all vendors, thus minimizing customization and services costs. LocalGrid's platform resides on devices between the edge and the cloud and provides reliable M2M communication between devices without having to go through the cloud. This allows applications to make real-time decisions right at the edge without having to deal with the high-latency of communicating with the cloud. Moreover, all LocalGrid devices can communicate with the cloud through open communication standards, realizing the concept of fog being an extension of cloud. Applications running on LocalGrid's platform can utilize the interplay between the fog and cloud to solve more complex problems.

LocalGrid's Fog Computing platform is shipped with LocalGrid vRTU, a software-based virtual remote terminal unit that transforms communications between edge devices into compatible open standards. vRTU can be installed on off-the-shelf as well custom solutions from OEMs, endowing devices with RTU capabilities and providing a single point for management of all the edge devices, thus cutting down customization and maintenance costs.

**ParStream**

ParStream is a real-time IoT analytics platform. Cisco and ParStream are working together to build a fast, reliable, and highly scalable infrastructure for analysis on the fog. Cisco is planning to use this infrastructure to enhance its current offerings and provide new types of services.

ParStream's offering of a Big Data Analytics Platform for IoT is contingent on its patented database technology ParStream DB. ParStream DB is a column-based In-memory database with a highly parallel and fault tolerant architecture which is built using patented indexing and compression algorithms. Being an in-memory database, it is ideal for fog devices- which typically limited disk space. ParStream can push down query execution to the edge where data is produced and perform analytics in a highly distributed fashion. Furthermore, ParStream has considerably small footprint, making it feasible to be deployed on embedded devices and fog-enabled devices like Cisco IOx.

**Prismtech Vortex**

VORTEX is a ubiquitous data sharing platform made for the Internet of Things. It provides scalable end-to-end seamless, efficient, secure and timely data sharing across IoT supporting devices, edges, gateways and cloud.

VORTEX leverages the DDS 2.0 standard for interoperable data sharing and extends it to support Internet Scale systems, mobility and Web 2.0 applications. VORTEX also seamlessly integrates with common IoT message passing protocols as MQTT and CoAP. In addition to address security and privacy requirements, VORTEX provides support for fine-grained access control and both symmetric and asymmetric authentication.

Each IoT device is connected to a Vortex edge device that executes all Vortex's software. Each of these softwares performs a function necessary for the realization of a globally shared DDS. A Vortex edge device with the IoT devices connected to it forms a domain (a DDS entity), called fog-domain in this context. Equipped with such devices, VORTEX supports a number of deployment models.

- Fog + Cloud: IoT devices inside a fog-domain communicate with each other in a peer-to-peer fashion. Those across fog-domains need to communicate through the cloud.

- Fog + Cloud-link + Cloud : Similar to the previous model, devices within the same fog-domain communicate peer-to-peer, while devices not in the same fog-domain exchange data through the cloud using a Cloud Link that handles the associated security issues and controls what information is exposed.

• Federated Fog: Each fog-domain has a Vortex Cloud-link running on the Vortex device. Federated Fog is a collection of fog domains, which are federated by Cloud-link instances. Information exchanged between fog-domains is controlled by Cloud-link instances.

## 4.8  Case Study

A smart city is one of the key use-cases of the Internet of Things, which in itself is a combination of a variety of use-cases ranging from smart traffic management to energy management of buildings. In this section, we present a case study on smart traffic management and show that employing fog computing improves the performance of the application in terms of response time and bandwidth consumption. A smart traffic management system can be realized by a set of stream queries executing on data generated by sensors deployed throughout the city. Typical examples of such queries are real-time calculation of congestion (for route planning), or detection of traffic incidents. In this case study, we compare the performance of a query DETECT_TRAFFIC_INCIDENT (as shown in Figure 4.5) on fog infrastructure versus the typical cloud implementation.

In the query, the sensors deployed on the roads send the speed of each crossing vehicle to the query processing engine. The operator "Average Speed Calculation" calculates the average speed of vehicles from the sensor readings over a given time frame and sends this information to the next operator. The operator "Congestion Calculation" calculates the level of congestion in each lane based on the average speed of vehicles in that lane. The operator "Incident Detection", based on the average level of congestion, detects whether an incident has occurred or not. This query was simulated on both fog-based as well as cloud-based stream query processing engine. The comparison of both strategies is presented in the next sections.

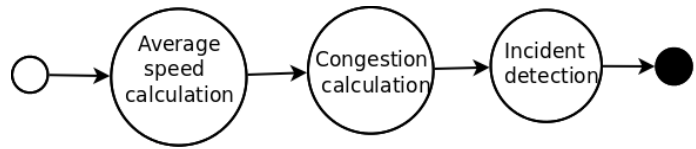

Figure 4.5 : Dag of query for incident detection

## 4.8.1 Experiment Setup

**Network Topology and Data Sources**

The network topology used for the simulation was a hierarchical topology of fog devices as described in [10]. The leaves of the tree-like topology are the edge devices (gateways) and the cloud is located at the root. Intermediate nodes in the tree represent intermediate network devices between the cloud and the edge - which are able to host applications by utilizing their nascent compute, network and storage capacity. Each fog device has an associated CPU capacity and an associated uplink network bandwidth, shall be utilized for running fog applications on them.

Traffic data fed to simulation was obtained from Sumo [19], a road traffic simulator. Induction loops were inserted on the road that measured the speed of vehicles, and the information was sent to the query processing engine.

The simulation environment was implemented in CloudSim [20] by extending the basic entities in the original simulator. Fog devices were realized by extending the Datacenter class, while stream operators were modeled as a VM in CloudSim. Furthermore, tuples that are executed by the stream operators were realized by extending Cloudlets. Fog devices are entities with only one host, whose resources it can provide for running applications. Each tuple has an associated CPU and network cost for processing it.

### 4.8.2 Performance Evaluation

**A) Average tuple delay**

Average tuple delay, as the name suggests, is the amount of time (on an average) that a tuple takes to be processed. Figure 4.6 compares the average end-to-end tuple delay experienced when running the query on fog against the case when a traditional cluster-based stream processing engine is used. The fog stream processing engine dynamically places operators across fog devices when there is enough capacity to save bandwidth and minimize latency. As Figure 4.6 shows, once operators placed on fog devices, the end-to-end tuple delay falls much below the delay of in-cloud processing as data are processed closer to the sources. However, it is worth mentioning that if the operators are not placed optimally, resource contention in edge devices can cause more delay.

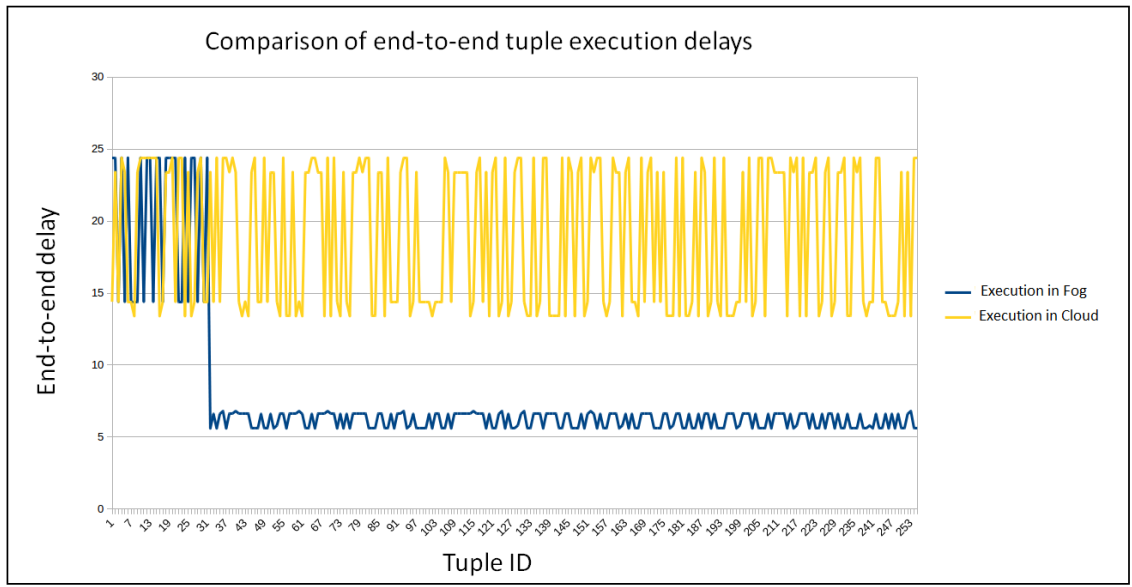

Figure 4.6 : Comparison of average end-to-end tuple execution delay

**b) Core Network usage**

In this experiment, we compare the core network usage for running the DETECT_TRAFFIC_INCIDENT query on fog-based and traditional cloud-based stream processing engine. Figure 4.7 shows that considerably less number of tuples traversing the core network once compared to the traditional cloud-based stream processing. Thus, running the query on the edge devices reduces the workload coming to the cloud for processing and also reduces the network usages considerably. However, as we discussed earlier, this reduction in network resource usage and end-to-end latency is only possible if a placement algorithm is in place to push operators downwards when enough capacity is available in edge devices.

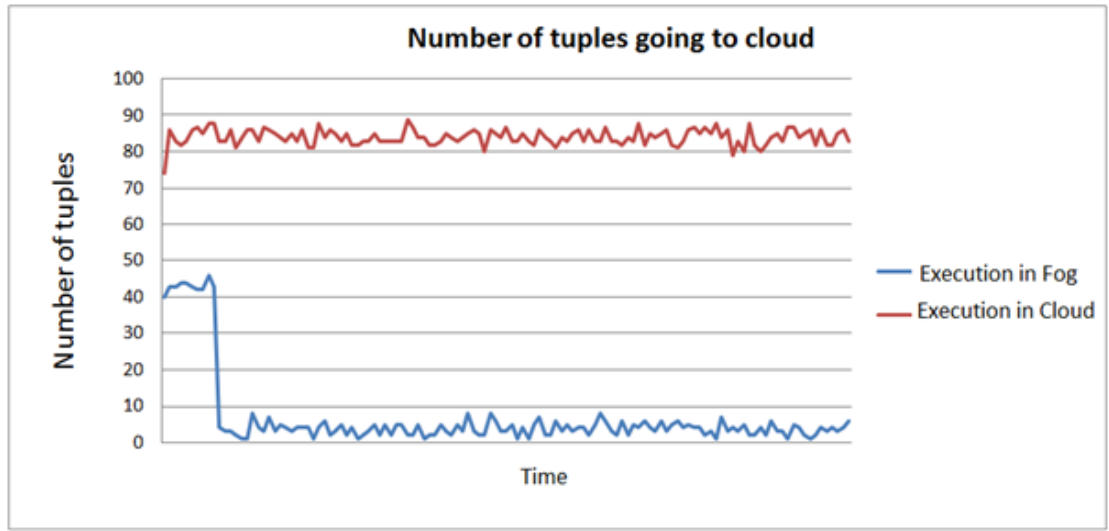

Figure 4.7 : Comparison of number of tuples reaching the cloud for processing, a measure of bandwidth consumption.

## 4.9 Summary

Fog computing is emerging as an attractive solution to the problem of data processing in the Internet of Things. It relies on devices on the edge of the network that have more p4rocessing power than the end devices and are nearer to these devices than the more powerful cloud resources, thus reducing latency for applications.

In this chapter, we introduced a reference architecture for IoT and discussed ongoing efforts in the academia and industry to enable the fog computing vision. Many challenges still remain though, with issues ranging from security to resource and energy usage minimization still in need for solutions. Open protocols and architectures are also other topics for future research that will make fog computing more attractive for end users.

# References

[1] Cao, Yu, et al. "FAST: A fog computing assisted distributed analytics system to monitor fall for stroke mitigation." Networking, Architecture and Storage (NAS), 2015 IEEE International Conference on. IEEE, 2015.

[2]  Stantchev, Vladimir, et al. "Smart Items, Fog and Cloud Computing as Enablers of Servitization in Healthcare." Sensors & Transducers (1726-5479)185.2 (2015).

[3]  Zhu, Jiang, et al. "Improving web sites performance using edge servers in fog computing architecture." Service Oriented System Engineering (SOSE), 2013 IEEE 7th International Symposium on. IEEE, 2013.

[4]  Aazam, Mohammad, and Eui-Nam Huh. "Fog Computing and Smart Gateway Based Communication for Cloud of Things." Future Internet of Things and Cloud (FiCloud), 2014 International Conference on. IEEE, 2014.

[5]  Zao, John K., et al. "Augmented brain computer interaction based on fog computing and linked data."Intelligent Environments (IE), 2014 International Conference on. IEEE, 2014.

[6]  Kosta, Sokol, et al. "Thinkair: Dynamic resource allocation and parallel execution in the cloud for mobile code offloading." INFOCOM, 2012 Proceedings IEEE. IEEE, 2012.

[7]  Chun, Byung-Gon, et al. "Clonecloud: elastic execution between mobile device and cloud." Proceedings of the sixth conference on Computer systems. ACM, 2011.

[8]  Banerjee, Arijit, et al. "MOCA: a lightweight mobile cloud offloading architecture." Proceedings of the eighth ACM international workshop on Mobility in the evolving internet architecture. ACM, 2013.

[9]  Orsini, Gabriel, Dirk Bade, and Winfried Lamersdorf. "Computing at the Mobile Edge: Designing Elastic Android Applications for Computation Offloading."

[10] Hong, Kirak, et al. "Mobile fog: A programming model for large-scale applications on the internet of things." Proceedings of the second ACM SIGCOMM workshop on Mobile cloud computing. ACM, 2013.

[11] Madsen, Henrik, et al. "Reliability in the utility computing era: Towards reliable Fog computing." Systems, Signals and Image Processing (IWSSIP), 2013 20th International Conference on. IEEE, 2013.

[12] Aazam, Mohammad, and Eui-Nam Huh. "Dynamic resource provisioning through Fog micro datacenter." Pervasive Computing and Communication Workshops (PerCom Workshops), 2015 IEEE International Conference on. IEEE, 2015.

[13] Satyanarayanan, M., Bahl, P., Cáceres, R., Davies, N. 2009. The Case for VM-Based Cloudlets in Mobile Computing. IEEE Pervasive Computing vol.8, no.4, 14–23.

[14] Lewis, Grace, et al. "Tactical cloudlets: Moving cloud computing to the edge." Military Communications Conference (MILCOM), 2014 IEEE. IEEE, 2014.

[15] Dsouza, Clinton, Gail-Joon Ahn, and Marthony Taguinod. "Policy-driven security management for fog computing: Preliminary framework and a case study." Information Reuse and Integration (IRI), 2014 IEEE 15th International Conference on. IEEE, 2014.

[16] Misra, Prasant, Yogesh Simmhan, and Jay Warrior. "Towards a Practical Architecture for the Next Generation Internet of Things." arXiv preprint arXiv:1502.00797 (2015).

[17] Kitchin, Rob. "The real-time city? Big data and smart urbanism." GeoJournal 79.1 (2014): 1-14.

[18] Cortés, Rudyar, et al. "Stream Processing of Healthcare Sensor Data: Studying User Traces to Identify Challenges from a Big Data Perspective." Procedia Computer Science 52 (2015): 1004-1009.

[19] Behrisch, Michael, et al. "SUMO–Simulation of Urban MObility." The Third International Conference on Advances in System Simulation (SIMUL 2011), Barcelona, Spain. 2011.

[20] Calheiros, Rodrigo N., et al. "CloudSim: a toolkit for modeling and simulation of cloud computing environments and evaluation of resource provisioning algorithms." Software: Practice and Experience 41.1 (2011): 23-50.